\documentclass[prd,aps,twocolumn,preprintnumbers,amsmath,amssymb,nofootinbib,superscriptaddress,notitlepage]{revtex4-1}

\usepackage{epsfig}
\usepackage[utf8]{inputenc}
\usepackage{color}
\usepackage{epstopdf}
\usepackage[colorlinks=true,
linkcolor=blue,
breaklinks=true,
urlcolor=blue,
citecolor=blue]{hyperref}

\newcommand{\nn}{\nonumber}
\newcommand{\be}{\begin{equation}}
\newcommand{\ee}{\end{equation}}
\newcommand{\bea}{\begin{eqnarray}}
\newcommand{\eea}{\end{eqnarray}}
\newcommand{\ba}{\begin{array}}
\newcommand{\ea}{\end{array}}
\newcommand{\bi}{\begin{itemize}}
\newcommand{\ei}{\end{itemize}}

\newcommand{\NKU}{\affiliation{ Nankai University, Tianjin 300000, China}}

\newcommand{\ucas}{\affiliation{University of Chinese Academy of Sciences, Beijing 100049, China}}

\newcommand{\keylab}{\affiliation{State Key Laboratory of Heavy Ion Science and Technology, Institute of Modern Physics, Chinese Academy of Sciences, Lanzhou 730000, China}}

\newcommand{\hebei}{\affiliation{Department of Physics, Hebei University, Baoding 071002, China}}

\begin{document}

\title{Entanglement redistribution of hyperon-antihyperon pair via sequential decay}

\author{Cong Li}\email{licong9048@163.com}
\NKU

\author{Xu Cao}\email{corresponding author: caoxu@impcas.ac.cn}
\keylab
\ucas

\author{Ai-Qiang Guo}\email{guoaq@impcas.ac.cn}
\keylab
\ucas

\author{Chun-Xu Yu}\email{yucx@nankai.edu.cn}
\NKU

\author{Hong-Wei Zhang}\email{corresponding author: hweiz0929@163.com }
\hebei
\keylab

\author{Zhe Zhang}\email{zhangzhe@impcas.ac.cn}
\keylab

\date{\today}

\begin{abstract}
  \rule{0ex}{3ex}
    Hyperon-antihyperon pairs produced in high energy electron-positron annihilation constitute a naturally spin-entangled system in the high energy regime.
    Recently, a probabilistic amplification of entanglement, termed autodistillation, has been found in the daughter baryon-antibaryon pairs from hyperon decay and is constrained by an upper boundary.
    This work demonstrates that the quantum entanglement in this process may be accompanied by a decrease, constrained by a lower boundary, but will not be completely lost.
    Thus, the entanglement of these systems undergoes redistribution within the phase space during the sequential decays of hyperons, highlighting an important role of hyperon polarization.
    By using the explicit spin density matrix of baryon pairs, it is also found that quantumness of the system characterized by quantum discord always have the possibility to increase during decay processes, even when entanglement evaluated by concurrence and negativity does not increase.
\end{abstract}

\maketitle

\section{Introduction} \label{sec:intro}

The fundamental distinction between quantum and classical correlations remains a cornerstone of quantum mechanics~\cite{Aspect:1981zz,Horodecki:1996nc,Horodecki:2009zz}. Quantum correlations, an essential concept for understanding quantum information processing~\cite{Modi:2012baj}, has been successfully applied to frontiers in optical lattices~\cite{Melikhov:1993mv,Bloch:2008zzb}, trapped ions~\cite{Blatt:2012chk, Leibfried:2003zz}, superconducting qubits~\cite{Devoret:2004srl, Clarke:2008ufm}, where high-fidelity entanglement generation and manipulation are now routine. In recent years, high-energy physics has emerged as a compelling arena for probing quantum correlations in relativistic regimes.
Notably, spin entanglement in top-quark pairs has been experimentally investigated at Large Hadron Collider~\cite{ATLAS:2023fsd}. Beyond top quarks, entanglement have been theoretically explored in a variety of collider processes involving leptons~\cite{Ehataht:2023zzt,Ma:2023yvd,LoChiatto:2024dmx}, massive gauge bosons~\cite{Aguilar-Saavedra:2022wam},
hyperons~\cite{Hao:2009kj,Wu:2024asu,Wu:2024mtj,Wu:2024asu,Hong:2025drg,Jaloum:2025bkx,Wu:2025dds,Pei:2026rlh,Pei:2025ito,Fabbrichesi:2024rec,Pei:2025yvr,Li:2008dk} and different quark flavors~\cite{Cheng:2025cuv,Afik:2025grr,Fucilla:2025kit,Zhang:2025ean}.
Much attention is paid to the hyperon–antihyperon system in charmonium decays, where hyperon decay serves as a self polarimeter~\cite{Tornqvist:1980af,Qian:2020ini,Shi:2019kjf,Khan:2020seu}, enabling the reconstruction of spin correlations from the decay products. Capitalizing on this advantage, the BESIII collaboration recently reported a clean probe of quantum entanglement~\cite{BESIII:2025vsr}.

The investigation of quantum entanglement in relativistic regimes has gained significant traction, particularly as a means to understand the strong force from alternative perspectives and to scrutinize the feasibility of Bell tests at high-energy colliders~\cite{Abel:1992kz,Dreiner:1992gt,Shi:2004yt,Shi:2019kjf,Li:2024luk,Low:2025aqq,Bechtle:2025ugc,Abel:2025skj,Ai:2025wnt,Pei:2026wfu}. As another central concept in the context of quantum information, entanglement distillation is the probabilistic conversion of a large ensemble of mixed entangled states into a smaller set of maximally entangled pure states \cite{Bennett:1995tk,Bennett:1995ra}. While conventional distillation relies on Local collective Operations and Classical Communication (LOCC), particle physics offers a unique parallel where the entanglement between daughter particles can spontaneously increase relative to the initial mother system. This phenomenon, termed entanglement autodistillation~\cite{Aguilar-Saavedra:2024fig}, arises because the decay process functions as a Stochastic LOCC (SLOCC) process~\cite{Bennett:2000fte}.
Recently, this idea has been applied to various collider processes, including top quark~\cite{Aguilar-Saavedra:2023hss,Aguilar-Saavedra:2024hwd}, muon~\cite{Aguilar-Saavedra:2023lwb}, and hyperon-antihyperon pairs~\cite{Feng:2025ryr}. However, as decay acts as a local manipulation within an expanding phase space, it may also inherently diminish the system's entanglement~\cite{Aguilar-Saavedra:2024fig,Feng:2025ryr}. While previous studies have largely focused on the amplification of entanglement, the potential for its reduction remains largely unquantified. This paper aims to fill that gap by numerically investigating the conditions under which entanglement is diminished or redistributed during sequential hyperon decays.


To address the characterization of the quantum correlations, one usually works within the widely used entanglement-separability dichotomy first formalized by Werner~\cite{Werner:1989zz}. Among the various measures of quantum correlations, entanglement --- an inseparable, strong form of correlation --- has been studied most extensively. Practical quantifiers are developed such as concurrence~\cite{Wootters:1997id} (based on entanglement of formation) and negativity~\cite{Vidal:2002zz} (based on the positive partial transpose criterion).
In the context of quantum information, “quantumness" is a heuristic term capturing how much a physical system's state deviates from classical descriptions. This deviation typically manifests through entanglement, superposition, or other intrinsic quantum features. While useful in qualitative discussions, precise analyses typically rely on well-defined measures like entanglement entropy or quantum discord. Discord quantifies the difference between two quantum generalizations of classical mutual information~\cite{Ollivier:2001fdq,Henderson:2001wrr}.
Crucially, it can capture non-classical correlations even in separable states that lack entanglement, providing a broader and more comprehensive framework for identifying and quantifying useful resources in quantum information and complex systems.

The study of quantum entanglement at high-energy colliders has a long history, dating back to analyses of entangled photons and neutral kaon pairs \cite{Shi:2025ggs}.
Today, using entanglement as both a test of Bell locality \cite{Aguilar-Saavedra:2026rsx,Low:2025aqq} and a probe of new physics \cite{Fabbrichesi:2022ovb} motivates a core and active research direction.
Produced via electron-positron annihilation, spin-entangled hyperon-antihyperon pair as massive qubit system is proposed as a unique platform to investigate the entanglement at hadronic scale \cite{Tornqvist:1980af}.
In particle physics, the angular distribution of decay products is leveraged as spin analyzers. The characteristic weak decays of hyperon (e.g., $\Xi^- \to \Lambda \pi^-$), yielding a final state consisting of a spin-1/2 baryon and a spinless pseudoscalar meson,
ensure that the complete spin information of the mother hyperon is cleanly transferred to the daughter baryon \cite{Lee:1957qs}.
By using the decay of daughter baryons as spin analyzers, their entanglement can be reconstructed as well.
Therefore a direct comparison of aforementioned entanglement measure of mother hyperon-antihyperon pairs and daughter baryon-antibaryon pairs allows for the investigation of entanglement evolution through sequential hyperon decays, where entanglement increase in certain decay chains is referred to as entanglement autodistillation ~\cite{Aguilar-Saavedra:2024fig,Feng:2025ryr}.
This paper pays special attention to the entanglement decrease arising from the local nature of the decay and the inherent enlargement of phase-space. Specifically, these factors drive the non-unitary evolution of the spin density matrix (SDM) and lead to an overall dilution of the quantum resources~\cite{Zurek:2003zz, Schlosshauer:2019ewh}.
Nevertheless, this evolution produces a distinct correlation profile in the daughter pairs, which serves as a sensitive diagnostic tool. This profile offers further insights into the underlying decay amplitudes and the efficiency of entanglement autodistillation within high-energy environments. The phase-space averaged profiles by an integration over helieicty angles of daughter baryons are also shown as an interesting comparison.

This paper is organized as follows. Sec.~\ref{sec:SDM} gives the spin density matrix of hyperon pairs and Sec.~\ref{sec:con_neg} presents the numerical results of concurrence, negativity, and discord. Sec.~\ref{sec:disc} discusses the underlying mechanism of entanglement redistribution and experimental prospects, and Sec.~\ref{sec:sum} briefly summarizes the results.

\section{Spin density matrix} \label{sec:SDM}

\begin{figure}[t]
    \centering
    \includegraphics[width=\linewidth]{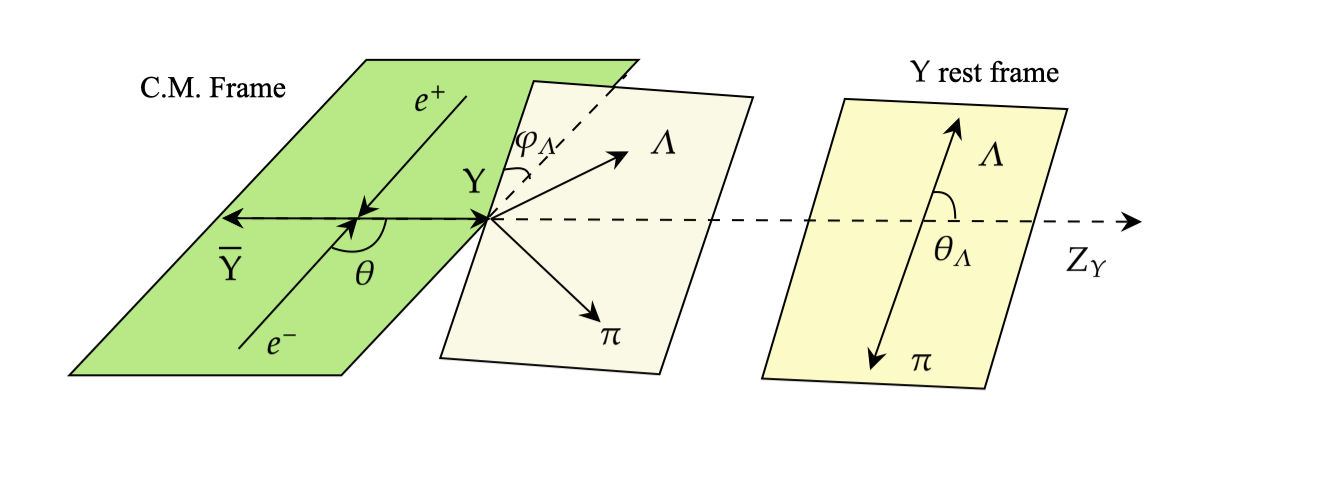}
    \caption{Definitions of the helicity angles for $e^+e^- \to Y \bar{Y}$ and the sequential decay $Y \rightarrow \Lambda \pi$. The helicity angles of $\bar Y \rightarrow \bar{\Lambda} \pi$ defined in the same manner are omitted.}
    \label{fig:helicity}
\end{figure}

This paper considers the $e^+e^- \to \psi \to Y \bar{Y}$ processes following the sequential decay of octet baryons $Y \rightarrow \Lambda \pi$ and $\bar Y \rightarrow \bar{\Lambda} \pi$, particularly for $Y = \Xi^-$ or $\Xi^0$.
The helicity framework is employed and can be easily extended to the annihilation of electron-positron to a pair of other hyperon-antihyperon through virtual photon or vector charmonium.
The helicity angles are constructed as illustrated in Fig.~\ref{fig:helicity}. The polar angle $\theta$ is the angle between the $Y$ momentum direction and the $e^+$ beam direction in the centre-of-mass (C.M.) frame of electron-positron system. The angles $\theta_{\Lambda}$ and $\varphi_{\Lambda}$ are the polar and azimuthal angles of the $\Lambda$ momentum direction in the $Y$ helicity system. In this system, the coordinate axis $z_{Y}$ is defined along the $Y$ momentum direction in the $e^{+}e^{-}$ C.M. frame, and the corresponding angles of the antiparticle decay sequence are obtained analogously.
The SDM of $Y \bar{Y}$ mother pairs is explicitly written down within the helicity formalism~\cite{Perotti:2018wxm,Salone:2022lpt,Cao:2024tvz,Zhang:2024rbl,Zhao:2025cbd,Zhang:2025oks,Zhao:2025cbd,Guo:2025bfn}:
\begin{widetext}
\begin{equation}
\label{eq:Theta_mu_nu_matrix}
\Theta_{\mu\nu}^{Y \bar{Y}} =
\begin{pmatrix}
 1+ \alpha_{\psi}\cos^2\theta & 0 & \beta_\psi \cos\theta \sin\theta & 0 \\
 0 & \sin^2\theta & 0 & \gamma_\psi \cos\theta \sin\theta \\
 -\beta_\psi \cos\theta \sin\theta & 0 & \alpha_\psi \sin^2\theta & 0 \\
 0 & -\gamma_\psi \cos\theta \sin\theta & 0 &-\alpha_\psi-\cos^2\theta
\end{pmatrix},
\end{equation}
\end{widetext}
with $\alpha_\psi \in [-1, +1]$ being the angular distribution parameter of the vector charmonium state and $\beta_\psi = \sqrt{1 - \alpha_\psi^2} \sin(\Delta\Phi)$,
$\gamma_\psi = \sqrt{1 - \alpha_\psi^2} \cos(\Delta\Phi)$.
Here $\Delta\Phi \in [-\pi, +\pi]$ is the relative phase between electric and magnetic form factor of $Y$.
The parameters for $J/\psi$ and $\psi(2S)$ are given in Table~\ref{tab:charmonium}.
The coordinate system $\{x_Y, y_Y, z_Y\}$ in the helicity system of $Y$ and $\bar{Y}$ is consistent with that in Refs.~\cite{Salone:2022lpt,Zhang:2023box}.
The entanglement measures are frame-independent and do not depend on the choice of other coordinates~\cite{Zhao:2025cbd}.
As can be seen, the produced hyperon is transversely polarized, in contrast to structureless particles, which exhibit null polarization.

With the help of the decay matrix of hyperons, the SDM for the daughter baryon-antibaryon pairs (herein $\Lambda \bar{\Lambda}$) can be directly calculated as \cite{Perotti:2018wxm,Salone:2022lpt,Zhang:2025oks,Zhang:2024rbl,Guo:2025bfn,Zhang:2023box}
\begin{equation}
\label{eq:S_mu_nu_U2_matrix}
\Theta_{\mu\nu}^{\Lambda \bar{\Lambda}} =
\begin{pmatrix}
\Theta_{11} & \Theta_{12} & \Theta_{13} & \Theta_{14} \\
\Theta_{21} & \Theta_{22} & \Theta_{23} & \Theta_{24} \\
\Theta_{31} & \Theta_{32} & \Theta_{33} & \Theta_{34} \\
\Theta_{41} & \Theta_{42} & \Theta_{43} & \Theta_{44}
\end{pmatrix},
\end{equation}
being positive semi-definite with $\Theta_{ij} = \Theta_{ji} (Y \leftrightarrow \bar{Y}, \Lambda \leftrightarrow \bar{\Lambda},  \theta_Y \leftrightarrow \pi - \theta_{\bar{Y}})$ when $i \neq j$.
The detailed expressions for the independent ten matrix elements in Eq.~\eqref{eq:S_mu_nu_U2_matrix} are given in Appendix~\ref{app:matrix} in terms of the decay parameters ($\alpha_{\Xi}$, $\phi_{\Xi}$) for the $\Xi \to \Lambda \pi$ as well as ($\alpha_{\bar \Xi}$, $\phi_{\bar \Xi}$) for its charge-conjugated mode $\bar \Xi \to \bar \Lambda \pi$ in Table~\ref{tab:decay}.
The related parameters $\beta_\Xi$ and $\gamma_\Xi$ are given by $\beta_\Xi = \sqrt{1 - \alpha_\Xi^2}\sin(\phi_{\Xi})$ and $\gamma_\Xi = \sqrt{1 - \alpha_\Xi^2} \cos(\phi_\Xi)$.
Note that relative big uncertainties of $\phi_{\Xi}$ and $\phi_{\bar \Xi}$ are present in different processes~\cite{BESIII:2021ypr,BESIII:2022lsz,BESIII:2023drj,BESIII:2025dke} but they do not influence the final results as discussed in Ref.~\cite{Feng:2025ryr} and Sec.~\ref{sec:disc}.

Those SDMs in Eqs.~\eqref{eq:Theta_mu_nu_matrix} and~\eqref{eq:S_mu_nu_U2_matrix}  are starting point to investigate the entanglement of mother and daughter systems.
If considering a pair of spin-half baryon-antibaryon as a massive two-qubit quantum system, the density matrix can be expressed as:
\begin{equation}
\label{eq:spin_density_matrix}
\begin{split}
\rho_{Y\bar{Y}} = \frac{1}{4} \bigl[
    & 1 \otimes 1 + \mathbf{P}^+ \cdot \mathbf{\sigma} \otimes 1 + 1 \otimes \mathbf{P}^- \cdot \mathbf{\sigma} \\
    & + \sum_{i,j} C_{ij} \, \sigma_i \otimes \sigma_j \bigr],
\end{split}
\end{equation}
with $\mathbf{\sigma} = (\sigma_1, \sigma_2, \sigma_3)$ being Pauli matrices, $\mathbf{P}^\pm$ is the polarization vectors of hyperon and antihyperon, and $C_{ij}$ is their polarization correlation matrix.
Eq.~\eqref{eq:spin_density_matrix} can be expressed more compactly with the form
$\rho_{Y\bar{Y}} = \frac{1}{4} \, \Theta_{\mu\nu} \, \sigma_\mu \otimes \sigma_\nu
\quad \text{with} \quad
\Theta_{00}=1,\;
\Theta_{i0}=P_i^+,\;
\Theta_{0j}=P_j^-,\;
\Theta_{ij}=C_{ij}.$
Here, $\sigma_0$ is defined as the $2\times2$ identity matrix.

\begin{table*}[hbt]
\vspace{-2mm}
\centering
\caption{The angular distribution parameter $\alpha_\psi$ and the relative phase $\Delta\Phi$ of $J/\psi$ and $\psi(2S) \to Y \bar{Y}$ for $Y = \Xi^0$ or $\Xi^-$. The first and second uncertainties are statistical and systematical, respectively.}
\label{tab:charmonium}
\begin{tabular}{lcc}
\hline\noalign{\smallskip}
Decay &  $\alpha_\psi$ & $\Delta\Phi /$rad \\
\hline
$J/\psi \to \Xi^0 \bar \Xi^0$  & 0.514$\pm$0.006$\pm$0.015~\cite{BESIII:2023drj}  & 1.168$\pm$0.019$\pm$0.018~\cite{BESIII:2023drj}  \\

$J/\psi \to \Xi^- \bar \Xi^+$  &
0.586$\pm$0.012$\pm$0.010~\cite{BESIII:2021ypr}  & 1.213$\pm$0.046$\pm$0.016~\cite{BESIII:2021ypr}  \\

\noalign{\smallskip}\hline\noalign{\smallskip}
$\psi(2S) \to \Xi^0 \bar \Xi^0$ &  0.768$\pm$0.029$\pm$0.025~\cite{BESIII:2025dke}  & 0.257$\pm$0.061$\pm$0.009~\cite{BESIII:2025dke}  \\

$\psi(2S) \to \Xi^- \bar \Xi^+$ &  0.693$\pm$0.048$\pm$0.049~\cite{BESIII:2022lsz}   & 0.667$\pm$0.111$\pm$0.058~\cite{BESIII:2022lsz} \\
\noalign{\smallskip}\hline\noalign{\smallskip}
\hline
\end{tabular}
\end{table*}


\begin{table}[hbt]
\vspace{-2mm}
\centering
\caption{The decay parameters for the $\Xi \to \Lambda \pi$ and $\bar \Xi \to \bar \Lambda \pi$ decay \cite{ParticleDataGroup:2024cfk}.}
\label{tab:decay}
\begin{tabular}{lcc}
\hline\noalign{\smallskip}
Decay &  $\alpha_Y$ & $\phi_Y $ \\
\hline
$\Xi^0 \to \Lambda \pi^0$  & -0.349$\pm$0.009   & 0.3$\pm$0.6$^{\circ}$ \\

$\bar{\Xi}^0 \to \bar{\Lambda} \pi^0$  & 0.379$\pm$0.004  & -0.3$\pm$0.6$^{\circ}$\\

\noalign{\smallskip}\hline\noalign{\smallskip}
 $\Xi^- \to \Lambda \pi^-$ &   -0.390$\pm$0.007  & -1.2$\pm$1.0$^{\circ}$\\

 $\bar{\Xi}^+ \to \bar{\Lambda} \pi^+$ &  0.371$\pm$0.007 &-1.2$\pm$1.2$^{\circ}$\\

\noalign{\smallskip}\hline\noalign{\smallskip}
\hline
\end{tabular}
\end{table}

\section{Entanglement Measures}
\label{sec:con_neg}

\subsection{Relations between entanglement measures and quantum discord}

Quantum entanglement is a quintessential feature of quantum mechanics that has no counterpart in classical physics \cite{Horodecki:2009zz,Brunner:2013est}. For mixed states of two qubits, concurrence and negativity serve as practical and widely adopted entanglement measures; the latter is also a monotone applicable to arbitrary dimensions of bipartite system.
For two-qubit systems, concurrence is directly related to the entanglement of formation, and negativity is derived from the Positive Partial Transpose (PPT) criterion for separability~\cite{Horodecki:1996nc}.
Together, they effectively characterize the entanglement resource of the system.

Based on the study of different measures of quantum information, entanglement does not capture all nonclassical correlations and that even separable states usually contain correlations, known as quantum discord, that are not entirely classical~\cite{Modi:2012baj}.
In this paper, concurrence $C(\rho)$, negativity $N(\rho)$, and quantum discord $D^{Y}[\rho]$ are used to aptly assess the evolution of quantum entanglement during the hyperon decay process.
Their values are bounded by $ 0 \le C(\rho) \le 1 $, $ 0 \le N(\rho) \le 1/2 $ and $ 0 \leq D^{Y}[\rho] \leq 1 $ as detailed in the following subsections.
Both concurrence and negativity indicate the separability of the system when they are zero, and the presence of entanglement whenever they are non-zero.
For two-qubit pure states or Bell diagonal states, concurrence and negativity are closely
related by relation $C(\rho) =2 N(\rho)$, but for mixed states they can differ.
Quantum discord can be non-zero even for separable states,
which explains the key observation in Sec. \ref{sec:discord}.

\subsection{Concurrence}

A state ($Y \bar{Y}$ here) is called separable if and only if its representing density matrix $\rho$ can be written as a linear convex combination of tensor products of density matrices for the subsystems, e.g. $\rho = \sum_{ij} p_{ij} \,\rho_i^{Y} \otimes \rho_j^{\bar{Y}}$, otherwise the state exhibiting quantum correlations are called entangled or non-separable~\cite{Werner:1989zz}.

The eigenvalues $\lambda_i$ of a generic density matrix $\rho$ satisfy $0 \leq \lambda_i \leq 1$ with $\sum_i |\lambda_i| = 1$ and Tr$[\rho^2] \leq 1$.
For pure states $\lambda_i$ are 1 (non-degenerate) and 0 ($n-1$ times degenerate), such that Tr$[\rho^2]$ reaches the upper bound.
The concurrence of the pure states in arbitrary dimensions is $C(\rho) = \sqrt{2(1-\mathrm{Tr}[(\rho_Y)^2])} = \sqrt{2(1-\mathrm{Tr}[(\rho_{\bar{Y}})^2])}$
 with $\rho_Y$ and $\rho_{\bar{Y}}$ being the reduced density matrices of the subsystems.
In principle, the concurrence of any mixed state $\rho$ can be defined by decomposing it into a set of pure states, but only for two dimensional systems it has a simple analytic solution
with respect to the associated spin-flipped density matrix $\sqrt{ \sqrt{\rho} (\sigma_y \otimes \sigma_y) \rho^* (\sigma_y \otimes \sigma_y) \sqrt{\rho} }$.
Here $\rho^*$ denotes the complex conjugate of the $\rho$ taking in the $\sigma_z$ eigenbasis.
The eigenvalues of above Hermitian matrix are denoted as $r_i$, $i = 1, 2, 3, 4$ in decreasing order.
The concurrence of the bipartite system is then defined as~\cite{Wootters:1997id}:
\begin{equation}
\label{eq:concurrence_definition}
C(\rho) \equiv \max\left(r_1 - r_2 - r_3 - r_4, 0\right),
\end{equation}
with $ 0 \le C(\rho) \le 1 $.
The system is separable if $C(\rho) = 0$ and otherwise it is entangled.
After some local unitary transformation, the SDM of $Y \bar{Y}$ system in Eq.~\eqref{eq:Theta_mu_nu_matrix} can be written as a symmetric $X$ state, the concurrence of which is of analytical form \cite{Wu:2024asu}.

\begin{figure}[t]
    \centering
    \includegraphics[width=0.9\linewidth]{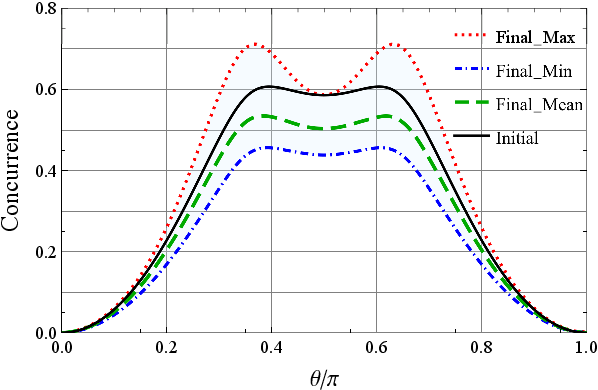}\put(-185,120){(\boldmath a)}

    \includegraphics[width=0.9\linewidth]{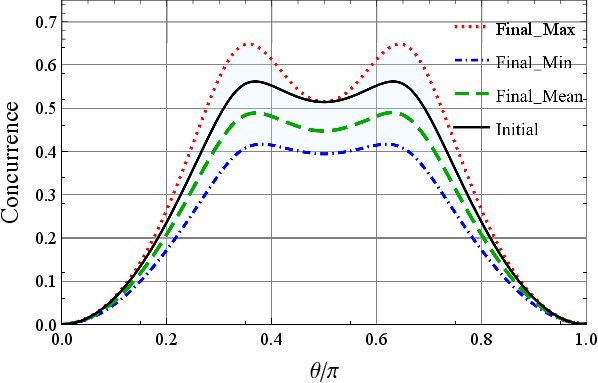}\put(-185,120){(\boldmath b)}

     \caption{Concurrence of $\Xi \bar{\Xi}$ (solid black line) and $\Lambda\bar{\Lambda}$ (light blue shaded area between red dotted line and blue dot-dashed line) in the
   (a) $e^+e^-\to J/\psi \to \Xi^{-} (\to \Lambda \pi^-)~\bar{\Xi}^{+} (\to \bar{\Lambda}\pi^+)$ and (b) $e^+e^-\to J/\psi \to \Xi^{0} (\to \Lambda \pi^0)~\bar{\Xi}^{0} (\to \bar{\Lambda}\pi^0)$ process. The green long-dashed line represents the phase-space-averaged values.}

    \label{fig:JpsiXi_con}
\end{figure}

Concurrence of $Y \bar{Y}$ and $\Lambda\bar{\Lambda}$ in the $e^+e^- \to \psi \to Y (\to \Lambda \pi)~\bar{Y} (\to \bar{\Lambda}\pi)$ process are shown as a function of scattering angle $\theta$ in Figs.~\ref{fig:JpsiXi_con} and~\ref{fig:psi2SXi_con} for $\psi = J/\psi$ or $\psi(2S)$ and $Y = \Xi^-$ or $\Xi^0$.
As noted in Appendix \ref{app:matrix}, besides depending on $\theta$ the $\Lambda\bar{\Lambda}$ concurrence is also a function of the helicity angles of both hyperons, denoted as $(\theta_\Lambda, \varphi_\Lambda)$ and  $(\theta_{\bar{\Lambda}}, \varphi_{\bar{\Lambda}})$, respectively. The labels Final\_Max (red dotted line) and Final\_Min (blue dot-dashed line) denote the upper and lower boundaries of the concurrence for the $\Lambda\bar{\Lambda}$ system obtained by scanning over the full phase space of these four angular variables at each fixed scattering angle $\theta$.
The Final\_Mean is the concurrence for the $\Lambda\bar{\Lambda}$ system obtained by an integration over the helicity angles $(\theta_\Lambda, \varphi_\Lambda)$ and  $(\theta_{\bar{\Lambda}}, \varphi_{\bar{\Lambda}})$, which can be viewed as phase-space-averaged values. The results of the black curves are consistent with those reported in Refs.~\cite{Fabbrichesi:2024rec,Wu:2024asu} and the red lines of $J/\psi \to \Xi^0 \bar \Xi^0$, $J/\psi \to \Xi^- \bar \Xi^+$ and $\psi(2S) \to \Xi^- \bar \Xi^+$ are compatible with those in Ref.~\cite{Feng:2025ryr}.
The light blue region between the red and blue lines reveals the entanglement redistribution across the phase space of daughter $\Lambda\bar{\Lambda}$:
whereas the mother $Y \bar{Y}$ system is characterized by a single concurrence value at each fixed $\theta$, the daughter system exhibits a range of concurrence values spanning the kinematically allowed region.
It is addressed that the light blue areas above black lines signify the increase of entanglement after hyperon decay, or the so called quantum entanglement autodistillation. It is significant for $J/\psi$ but less pronounced for $e^+e^-\to \psi(2S) \to \Xi^{-} (\to \Lambda \pi^-)~ \bar{\Xi}^{+} (\to \bar{\Lambda}\pi^+)$, and even disappear for $e^+e^-\to \psi(2S) \to \Xi^{0} (\to \Lambda \pi^0)~\bar{\Xi}^{0} (\to \bar{\Lambda}\pi^0)$.
The light blue areas below black lines indicate the probabilistic reduction of entanglement between mother and daughter systems.
The autodistillation fails completely in $e^+e^-\to \psi(2S) \to \Xi^{0} (\to \Lambda \pi^0)~\bar{\Xi}^{0} (\to \bar{\Lambda}\pi^0)$ process, as quantified by the concurrence.
In all cases, the phase-space-averaged concurrence stay closer to their respective lower boundary, indicating that entanglement autodistillation does not occur for these averaged quantities.

The mother system $Y\bar{Y}$ is in a separable mixed state without entanglement when $\theta = 0$ and $\pi$, at which angles
the daughter state of $\Lambda\bar{\Lambda}$ is also separable without entanglement redistribution.
The entanglement has no possibility to increase during decay prosess but 	
necessarily decreases when $\theta = \pi/2$.
It is noted that mother system $Y\bar{Y}$ has neither polarization nor non-diagonal correlations at $\theta = 0$, $\pi/2$ and $\pi$ as seen in Eq.~\eqref{eq:Theta_mu_nu_matrix}.

The $\theta$-dependence of concurrence exhibits different patterns for the $J/\psi$ and $\psi(2S)$.
The largest concurrence occurs at $\theta = \pi/2$ for $\psi(2S)$, and the concurrence for the $J/\psi$ has two local maximum at certain angles $\theta_{max}$ symmetric with respect to $\pi/2$. The underlying mechanism has investigated in literature~\cite{Wu:2024asu} so will not discussed herein. Compared to the initial value, the peak values $\theta_{max}$ for both Final\_Max and Final\_Min moves a bit from 69.1$^{\circ}$ (or $65.5^\circ$) to 69.8$^{\circ}$ (or 69.8$^{\circ}$) and 66.1$^{\circ}$ (or 62.5$^{\circ}$) for $\Xi^{-}\bar{\Xi}^{+}$ (or $\Xi^{0}\bar{\Xi}^{0}$) respectively.
These scattering angles do not align with the electron beam direction, highlighting the potential for experimental observation of entanglement redistribution.
In Sec. IV, the $(\theta_\Lambda, \varphi_\Lambda)$ and  $(\theta_{\bar{\Lambda}}, \varphi_{\bar{\Lambda}})$ at $\theta_{max}$ will be further examined to confirm that they all fall within the acceptance of the central detector.

\begin{figure}[t]
    \centering
    \includegraphics[width=0.9\linewidth]{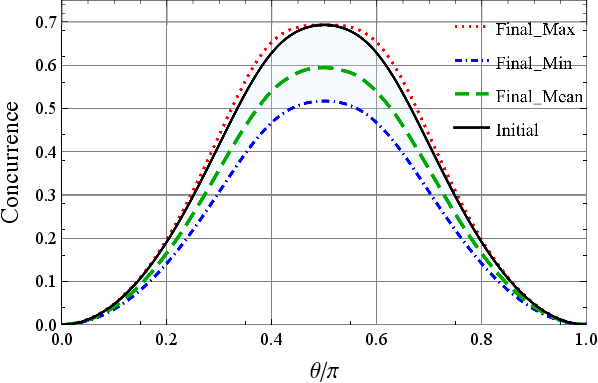}\put(-185,120){(\boldmath a)}

    \includegraphics[width=0.9\linewidth]{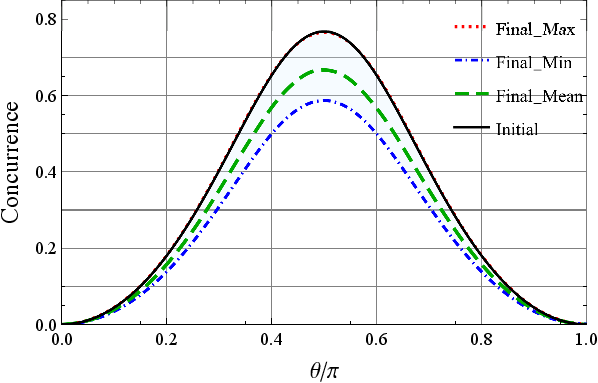}\put(-185,120){(\boldmath b)}

    \caption{Concurrence of $\Xi \bar{\Xi}$ (solid black line) and $\Lambda\bar{\Lambda}$ (light blue shaded area between red dotted line and blue dot-dashed line) in the
   (a) $e^+e^-\to \psi(2S) \to \Xi^{-} (\to \Lambda \pi^-)~\bar{\Xi}^{+} (\to \bar{\Lambda}\pi^+)$ and (b) $e^+e^-\to \psi(2S) \to \Xi^{0} (\to \Lambda \pi^0)~\bar{\Xi}^{0} (\to \bar{\Lambda}\pi^0)$ process.
   Note that the black and red curves are nearly superimposed, rendering them visually indistinguishable in (b). The green long-dashed line represents the phase-space-averaged values.
    }

    \label{fig:psi2SXi_con}
\end{figure}

\subsection{Negativity}

The bipartite system ($Y \bar{Y}$ here) is entangled if it does not remain positive under partial transposition with respect to any subsystem, as claimed by Peres-Horodecki criterion.
Negativity as another entanglement measure is defined in terms of the absolute sum of the eigenvalues $\kappa_i$ of the partially transposed density matrix $\rho^{T_Y}$ or $\rho^{T_{\bar Y}}$:
\begin{equation}
\label{eq:negativity_eigenvalues}
N(\rho) = \frac{\|\rho^{T_Y}\| - 1}{2} = \sum_i \frac{ |\kappa_i| - \kappa_i}{2}.
\end{equation}
where $\|\rho^{T_Y}\|$ is the trace norm or the sum of the singular values of $\rho^{T_Y}$.
For separable states the eigenvalues $\kappa_i$ are non-negative, leading to vanishing $N(\rho)$, since their $\rho^{T_Y}$ or $\rho^{T_{\bar Y}}$ remains positive semidefinite.
For entangled states, the appearance of negative eigenvalues forces \(\sum_i |\kappa_i| > 1\) so negativity quantifies the entanglement content of system with $ 0 \le N(\rho) \le 1/2 $.

\begin{figure}[t]
    \centering
    \includegraphics[width=0.9\linewidth]{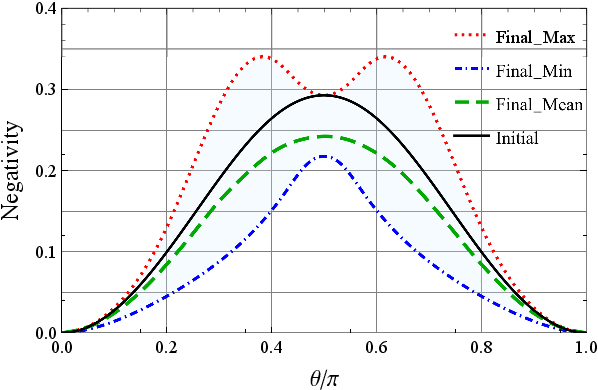}\put(-185,120){(\boldmath a)}

    \includegraphics[width=0.9\linewidth]{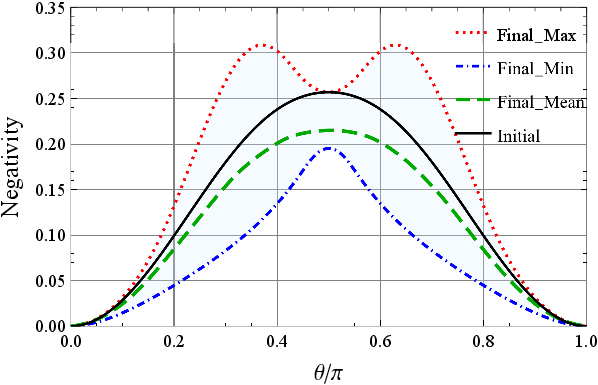}\put(-185,120){(\boldmath b)}

    \caption{Negativity of $\Xi \bar{\Xi}$ (solid black line) and $\Lambda\bar{\Lambda}$ (light blue shaded area between red dotted line and blue dot-dashed line) in the
   (a) $e^+e^-\to J/\psi \to \Xi^{-} (\to \Lambda \pi^-)~\bar{\Xi}^{+} (\to \bar{\Lambda}\pi^+)$ and (b) $e^+e^-\to J/\psi \to \Xi^{0} (\to \Lambda \pi^0)~\bar{\Xi}^{0} (\to \bar{\Lambda}\pi^0)$ process. The green long-dashed line represents the phase-space-averaged values.
    }

    \label{fig:JpsiXi_neg}
\end{figure}

\begin{figure}[t]
    \centering
    \includegraphics[width=0.9\linewidth]{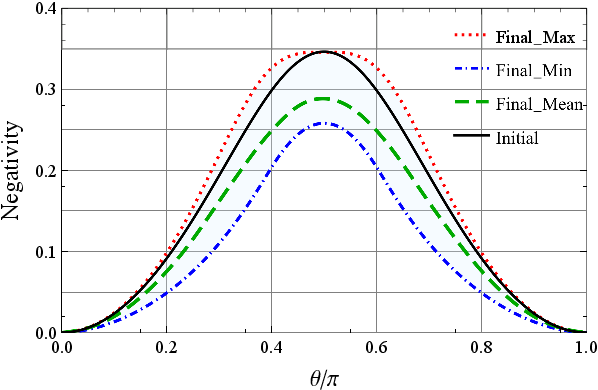}\put(-185,120){(\boldmath a)}

    \includegraphics[width=0.9\linewidth]{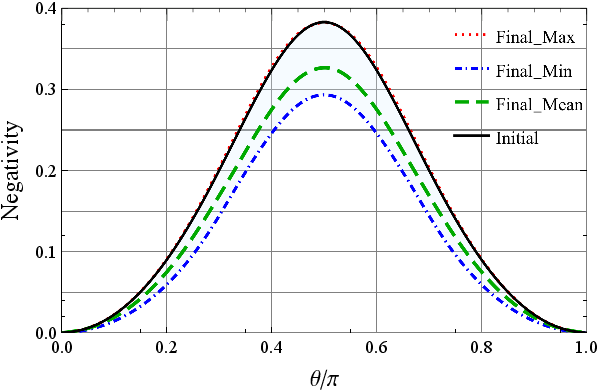}\put(-185,120){(\boldmath b)}
   \caption{Negativity of $\Xi \bar{\Xi}$ (solid black line) and $\Lambda\bar{\Lambda}$ (light blue shaded area between red dotted line and blue dot-dashed line) in the
   (a) $e^+e^-\to \psi(2S) \to \Xi^{-} (\to \Lambda \pi^-)~\bar{\Xi}^{+} (\to \bar{\Lambda}\pi^+)$ and (b) $e^+e^-\to \psi(2S) \to \Xi^{0} (\to \Lambda \pi^0)~\bar{\Xi}^{0} (\to \bar{\Lambda}\pi^0)$ process. Note that the black and red curves are nearly superimposed, rendering them visually indistinguishable in (b). The green long-dashed line represents the phase-space-averaged values.
    }
    \label{fig:psi2SXi_neg}
\end{figure}

Negativity of $Y \bar{Y}$ and $\Lambda\bar{\Lambda}$ for $e^+e^- \to \psi \to Y (\to \Lambda \pi^-)~\bar{Y} (\to \bar{\Lambda}\pi^+)$ process are shown as a function of $\theta$ in Figs.~\ref{fig:JpsiXi_neg} and~\ref{fig:psi2SXi_neg}, the latter of which ($\Lambda\bar{\Lambda}$) are optimized or integrated with respect to helicity angles $(\theta_\Lambda, \varphi_\Lambda)$ and $(\theta_{\bar{\Lambda}}, \varphi_{\bar{\Lambda}})$.
The main conclusion is quite similar to those as
evaluated by the concurrence.
Negativity of $\Lambda\bar{\Lambda}$ coming from the  $Y \bar{Y}$ decay is redistributed over the four-body phase space:
it increases in part of the phase space and decreases in the rest as shown by the light blue areas between red and blue curves. Entanglement autodistillation fails for phase-space-averaged negativity since the green dashed curves labeled as Final\_Mean remain closer to their respective lower boundary.

It can be seen the profiles of lower and upper boundary of daughter system are similar to those of mother system both for negativity and concurrence (and also discord in Sec.~\ref{sec:discord}), with an except of negativity in $e^+e^- \to J/\psi \to Y (\to \Lambda \pi^-)~\bar{Y} (\to \bar{\Lambda}\pi^+)$ process in Fig.~\ref{fig:JpsiXi_neg}.
The upper boundary of negativity in this process has two local maximum at certain angles $\theta_{max}= 69.8^{\circ}$ for  $\Xi^{-}\bar{\Xi}^{+}$ (or $66.1^\circ$ for $\Xi^{0}\bar{\Xi}^{0}$) symmetric with respect to $\pi/2$, and that of mother system $Y\bar{Y}$ reaches the peak at $\theta = \pi/2$.

Maximal enlargement is not achieved across full angles according to the calculation of negativity and concurrence of $Y \bar{Y}$ and $\Lambda\bar{\Lambda}$.
It is interesting that the $\eta_c$ and $\chi_{c0}$ decays into hyperon-antihyperon pairs exhibit maximal entangled due to the spinless nature of pseudoscalar charmonium ~\cite{Fabbrichesi:2024rec,Hong:2025drg}.
So that entanglement of daughter system e.g.~$p\bar{p}$ in the $e^+e^-\to \gamma \eta_c (\textrm{or } \chi_{c0}) \to \gamma \Lambda (\to p \pi^-) ~\bar{\Lambda} (\to \bar{p}\pi^+)$ process is postulated to always decrease in comparison of mother system $\Lambda\bar{\Lambda}$, a phenomenon similar to the $e^+e^-\to \psi(2S)\to \Xi^{0} (\to \Lambda \pi^0)~\bar{\Xi}^{0} (\to \bar{\Lambda}\pi^0)$ process.

The daughter systems $\Lambda\bar{\Lambda}$ are non-separable except at $\theta = 0$ and $\pi$ as shown by their lower boundary of negativity and concurrence.
Such that the entanglement may decrease but will not be completely lost during the hyperon decay processes.

\subsection{Quantum Discord} \label{sec:discord}


Quantum discord coincide with entanglement for pure states. It is more general than entanglement considering that any entangled state shows discord but the opposite is not true~\cite{Adesso:2016pfx}.
The definition of quantum discord requires the definition of the mutual information of a bipartite system:
$I(\rho) \equiv S(\rho_A) + S(\rho_B) - S(\rho) = S(\rho_A) - S(\rho_{A|B})$.
Here $S(\rho) \equiv -\operatorname{Tr}(\rho \log_{2} \rho)$ denotes the von Neumann entropy~\cite{Vedral:2002zz,Adesso:2016pfx}, and $S(\rho_{A|B}) \equiv S(\rho_{B}) - S(\rho_{AB})$ is the conditional entropy without measurement.
The $\rho_A = \operatorname{Tr}_B(\rho_{AB})$ and $\rho_B = \operatorname{Tr}_A(\rho_{AB})$ are the reduced density matrices of the subsystems.
In other terms, mutual information quantifies the amount of total correlations contained in system.

The amount of classical correlations can be obtained through local von Neumann measurements $\Pi^B$ performed on subsystem $B$.
Conditioned on this projective measurement 
the post-measurement state of subsystem $A$ is described by an ensemble
\begin{equation}
    \rho_{A|\Pi_k^B} = \operatorname{Tr}_{B} \left[ \frac{1}{{p_k}} (1^A \otimes \Pi_k^B) \rho_{AB} (1^A \otimes \Pi_k^B) \right],
\end{equation}
with the probability $p_k = \operatorname{Tr} \left[ (1^A \otimes \Pi_k^B) \rho_{AB}(1^A \otimes \Pi_k^B) \right]$ and $1^A$ being the identity operator for subsystem $A$.
For a qubit system in Eq.~\eqref{eq:spin_density_matrix}, $k = \pm$ and $p_\pm = (1 \pm \mathbf{\hat{n}} \cdot \mathbf{P}^-)/2$ with $\mathbf{\hat{n}}$ corresponding to the axis along which the spin of $B= \bar{Y}$ is measured.
The corresponding classical mutual information is
\begin{equation} \label{eq:clasinfor}
    J^A(\rho) \equiv S(\rho_A) - \sum_k p_k S(\rho_{A|\Pi_k^B}).
\end{equation}
A measurement-independent quantification of classical correlations is obtained by maximizing over all local von Neumann measurements on subsystem $B$, labeled as $\max_{\{\Pi_k^B\}}$.

Discord is defined as the difference between the amount of total correlations and the one of classical correlations:
\begin{equation}
        D^A[\rho]  \equiv I(\rho) - \max_{\{\Pi_k^B\}} J^A(\rho),
\end{equation}
with $ 0 \leq D^A[\rho] \leq 1 $ as a measure of the content of non-classical correlations of a bipartite system.
In this paper $D^{Y}[\rho] = D^{\bar{Y}}[\rho]$ since $CP$ violation is not enforced.
For a symmetric $X$ state by local unitary transforming on the $Y \bar{Y}$ system in Eq.~\eqref{eq:Theta_mu_nu_matrix}, the discord is of analytical form~\cite{Luo:2008ecu,Afik:2022dgh,Han:2024ugl,Wu:2025dds}.

\begin{figure}[t]
    \centering
    \includegraphics[width=0.9\linewidth]{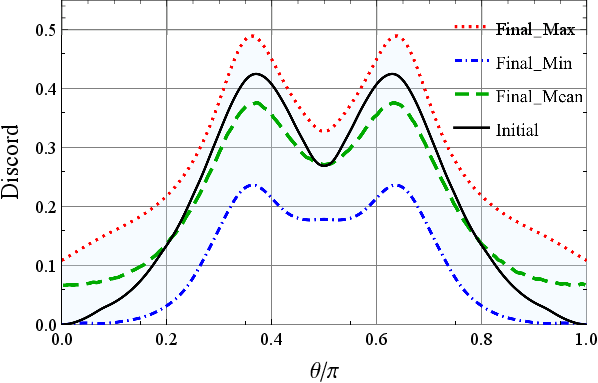} \put(-185,120){(\boldmath a)}

   \includegraphics[width=0.9\linewidth]{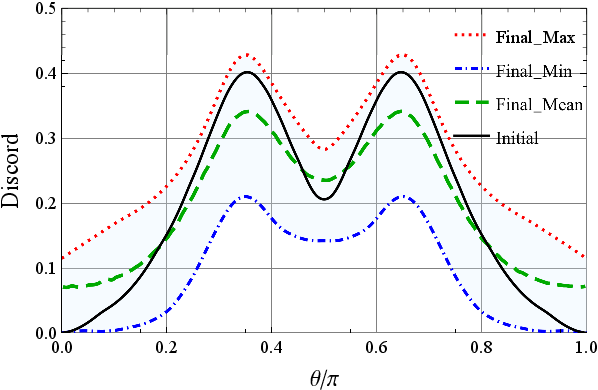} \put(-185,120){(\boldmath b)}

     \caption{Discord of $\Xi \bar{\Xi}$ (solid black line) and $\Lambda\bar{\Lambda}$ (light blue shaded area between red dotted line and blue dot-dashed line) in the
   (a) $e^+e^-\to J/\psi \to \Xi^{-} (\to \Lambda \pi^-)~\bar{\Xi}^{+} (\to \bar{\Lambda}\pi^+)$ and (b) $e^+e^-\to J/\psi \to \Xi^{0} (\to \Lambda \pi^0)~\bar{\Xi}^{0} (\to \bar{\Lambda}\pi^0)$ process. The green long-dashed line represents the phase-space-averaged values.
    }

    \label{fig:JpsiXi_discord}
\end{figure}

\begin{figure}[t]
    \centering
    \includegraphics[width=0.9\linewidth]{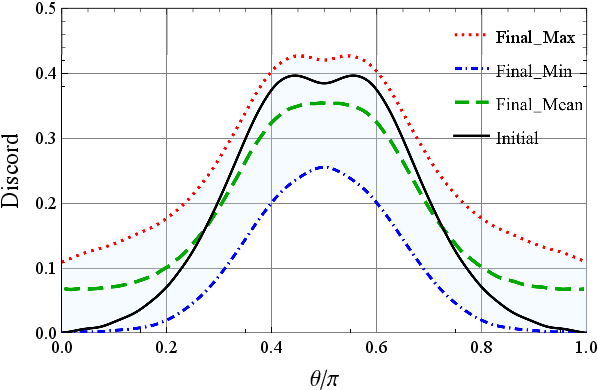}\put(-185,120){(\boldmath a)}

    \includegraphics[width=0.9\linewidth]{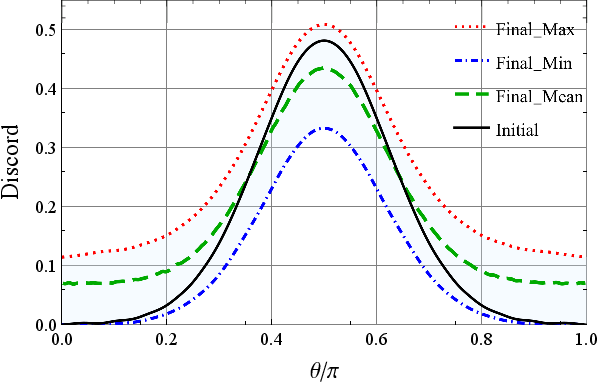} \put(-185,120){(\boldmath b)}

   \caption{Discord of $\Xi \bar{\Xi}$ (solid black line) and $\Lambda\bar{\Lambda}$ (light blue shaded area between red dotted line and blue dot-dashed line) in the
   (a) $e^+e^-\to \psi(2S) \to \Xi^{-} (\to \Lambda \pi^-)~\bar{\Xi}^{+} (\to \bar{\Lambda}\pi^+)$ and (b) $e^+e^-\to \psi(2S) \to \Xi^{0} (\to \Lambda \pi^0)~\bar{\Xi}^{0} (\to \bar{\Lambda}\pi^0)$ process. The green long-dashed line represents the phase-space-averaged values.
    }

    \label{fig:psi2SXi_discord}
\end{figure}

Discord of $Y \bar{Y}$ and $\Lambda\bar{\Lambda}$ for $e^+e^-\to \psi \to Y (\to \Lambda \pi^-)~\bar{Y} (\to \bar{\Lambda}\pi^+)$ process are shown as a function of $\theta$ in Figs.~\ref{fig:JpsiXi_discord} and~\ref{fig:psi2SXi_discord}.
The former ($Y \bar{Y}$) is consistent with the results reported in Ref.~\cite{Wu:2025dds}, and the latter ($\Lambda\bar{\Lambda}$) are appropriately optimized (or integrated) with respect to helicity angles of $\Lambda$ and $\bar{\Lambda}$ to identify the extreme values (or phase-space-averaged values).
A prominent feature is that the discord of daughter system has the possibility to increases across the entire range of scattering angles for all channels in comparison of the mother system when evaluated against the upper and lower boundaries.
As mentioned the entanglement of $\Lambda\bar{\Lambda}$ system is impossible to increase at $\theta = 0$, $\pi/2$ and $\pi$, but the discord has the possibility to be enhanced.
So that the quantumness of the system is increased even the entanglement is not enhanced in certain phase space.
This phenomenon happens across full phase space for $e^+e^-\to \psi(2S) \to \Xi^{0} (\to \Lambda \pi^0)~\bar{\Xi}^{0} (\to \bar{\Lambda}\pi^0)$ as shown in Fig.~\ref{fig:psi2SXi_discord}(b), in which process the entanglement of $\Lambda\bar{\Lambda}$ system does not increase at all.
At $\theta = 0$ and $\pi$ the states with non-zero quantum discord but null entanglement is possibly emergent.
The phase-space-averaged discord (green dashed lines) is under redistribution with respect to the scattering angle. It decreases along the beam direction ($\theta = 0$ and $\pi$) whereas it increases at other direction.

It is noted that their quantumness of these state manifests in the irreversibility of local measurements during correlation extraction --- a feature distinct from entanglement, which characterizes the intrinsic nonseparability of the quantum state itself.
Therefore the non-classical correlations always have the possibility to increase during decay processes.
As another prominent feature, the discord of $\Lambda\bar{\Lambda}$ decreases to be close to zero at some domain approaching $\theta = 0$ and $\pi$, indicating few content of non-classical correlations under certain phase space.

It is pointed out that, for their respective initial values, the $\Xi^{-}\bar{\Xi}^{+}$ and $\Xi^{0}\bar{\Xi}^{0}$ systems in the $J/\psi$ channel (shown in Fig.~\ref{fig:JpsiXi_discord}) achieve their maximum discord at pairs of angles symmetric about $\pi/2$, with the smaller angles given by $\theta_{\text{max}} = 65.5^\circ$ and $63.7^\circ$, respectively (see also Ref.~\cite{Wu:2025dds}). The extreme boundary of the daughter $\Lambda\bar{\Lambda}$ system reaches its maximum discord at $\theta_{\text{max}} = 63.8^\circ$, which differs insignificantly from that of the mother system.

\begin{figure}[t]
    \centering
    \includegraphics[width=0.9\linewidth]{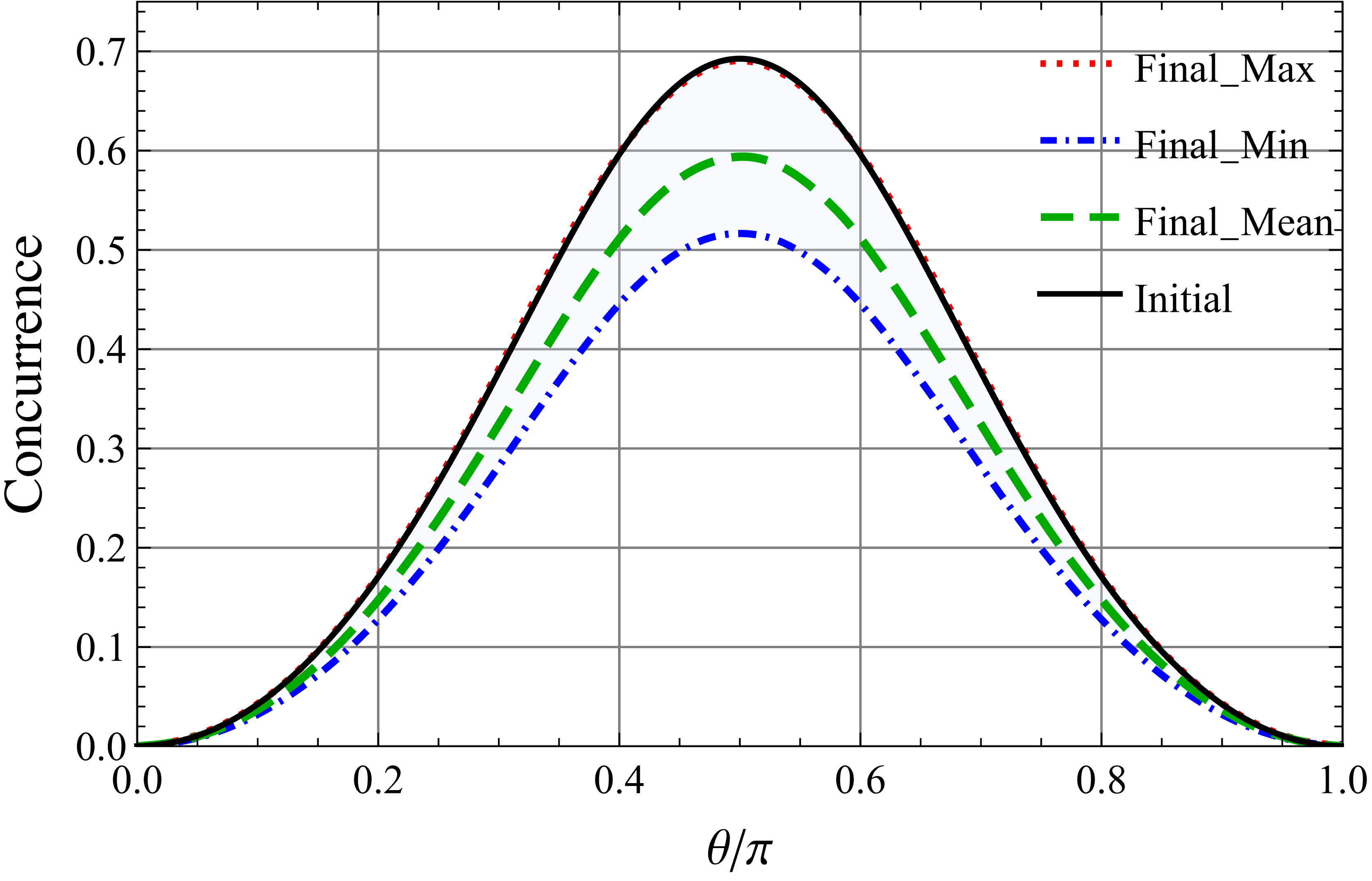}\put(-185,120){(\boldmath a)}

     \includegraphics[width=0.9\linewidth]{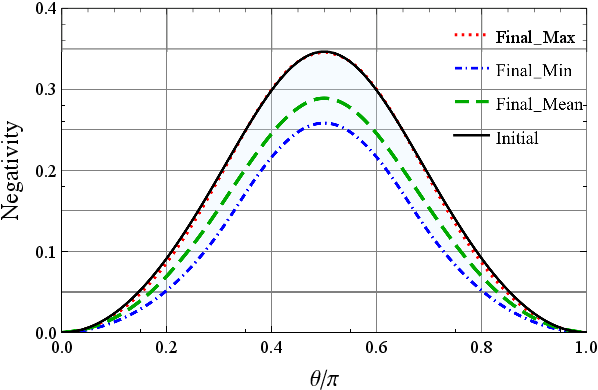}\put(-185,120){(\boldmath b)}

    \includegraphics[width=0.9\linewidth]{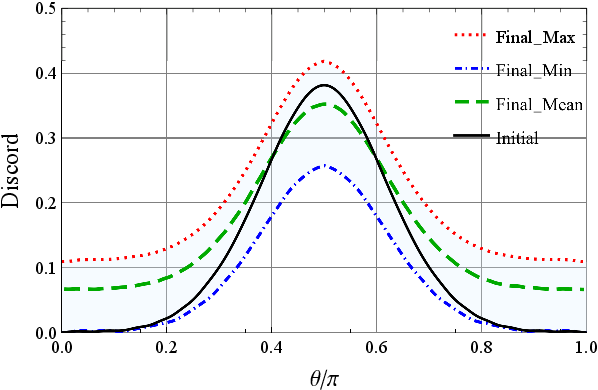}\put(-185,120){(\boldmath c)}

    \caption{(a) Concurrence and (b) negativity for $\Xi^{-}\bar{\Xi}^{+}$ (solid black line) and $\Lambda\bar{\Lambda}$ (light blue shaded area between the red dotted and blue dot-dashed lines) in the $e^+e^-\to \psi(2S) \to \Xi^{-} \bar{\Xi}^{+}$ process. For these results, $\Delta\Phi$ is set to 0.3 rad. Black and red curves are nearly identical and thus visually indistinguishable. (c) Quantum discord for the case where $\Delta\Phi = 0$ rad. The green long-dashed line represents the phase-space-averaged values.}

    \label{fig:psi2SXi_test}
\end{figure}


\section{Discussions and experimental prospects} \label{sec:disc}

The underlying mechanism responsible for the spontaneous entanglement change after the particles decay is attributed to the weak decay of hyperons as explained in Ref.~\cite{Feng:2025ryr}.
The non-zero decay parameter $\alpha_{Y}$ for $Y = \Xi^-$ or $\Xi^0$ in Table~\ref{tab:decay} comes from the parity violated amplitude of weak hadronic decay of hyperons.
Therefore the decay process as a local operator is not a unitary transformation, which
no longer maps two orthogonal polarization states of the mother particles to two orthogonal states of the daughter particles.
As a result, the entanglement of the decay product would be different from the mother system, probabilistic increase, or decrease as shown by the numerical calculation of the evolution of concurrence, negativity, and discord.
On the other hand, the decay phase $\phi_Y$ introduces only a phase
difference between the two spin states and it as a local unitary transformation does not change the entanglement of quantum systems during the decay process. This explains its vanishing impact on the upper and lower boundary of concurrence, negativity, and discord.

A mysterious discovery is that concurrence and negativity for $e^+e^-\to \psi(2S) \to \Xi^{0} (\to \Lambda \pi^0)~\bar{\Xi}^{0} (\to \bar{\Lambda}\pi^0)$ process are not enhanced at all when evaluated by the extreme values and phase-space-averaged quantities, though the non-classic correlations evaluated by discord has the possibility to increase.
On the other hand, entanglement in the $e^+e^-\to \psi(2S) \to \Xi^{-} (\to \Lambda \pi^-)~\bar{\Xi}^{+} (\to \bar{\Lambda}\pi^+)$ process increases insignificantly, hinting at the key role of relative phase $\Delta\Phi$ between electric and magnetic form factor of $\Xi$.
By adjusting the $\Delta\Phi$ to unphysicsl values, it is found that possibility of entanglement autodistillation of bipartite system in this channel is impossible to increase below a critical $\Delta\Phi = 0.3$ rad across the full phase space as shown in Fig.~\ref{fig:psi2SXi_test}(a,b).
Since small $\Delta\Phi$ means small transverse polarization of hyperon, it is concluded that transverse polarization of hyperon shall be big enough in order to generate the possibility of entanglement autodistillation in these processes.

It is also noted that the upper boundary of daughter system goes never down to the corresponding concurrence and negativity of the mother system.
The discord of daughter system always has the
possibility to increase during decay processes, even the hyperons produced in the electron–positron annihilation are unpolarized, as shown by the upper boundary and phase-space-averaged values
in Fig.~\ref{fig:psi2SXi_test}(c).

For experimental prospects, BESIII experiment has already accumulated a world-leading sample of 10 billion $J/\psi$ events and 2.7 billion $\psi(2S)$ events. As demonstrated in recent hyperon studies of the multi-dimensional angular analysis~\cite{BESIII:2021ypr, BESIII:2023drj, BESIII:2025dke}, these datasets provide sufficient statistics to determine the elements of spin density matrix with high precision.
It is feasible to experimentally observe entanglement redistribution and autodistillation in $e^+e^- \to J/\psi \to \Xi (\to \Lambda \pi)~\bar{\Xi} (\to \bar{\Lambda}\pi)$ process by using extreme values and phase-space-averaged quantities as tools.
Looking forward, the proposed STCF is expected to deliver a luminosity two orders of magnitude higher than BESIII, potentially collecting trillions of $J/\psi$ events~\cite{Achasov:2023gey}.
Such an unprecedented dataset would enable a precise probe of entangled hyperon systems and the more subtle manifestations of quantum discord in particle physics, potentially offering deeper insights into the strong interaction in the non-perturbative regime.

The boundary extrema --- whether maxima or minima --- of the concurrence, negativity, and quantum discord observed in the daughter systems arise purely from kinematic constraints within the available phase space. Table~\ref{tab:upper_xi} and~\ref{tab:lower_xi} give the angles $\theta_{max}$, $(\theta_\Lambda, \varphi_\Lambda)$ and $(\theta_{\bar{\Lambda}}, \varphi_{\bar{\Lambda}})$ that realize peak values of concurrence, negativity, and quantum discord for $e^+e^-\to J/\psi \to \Xi (\to \Lambda \pi)~\bar{\Xi} (\to \bar{\Lambda}\pi)$ process.
Those values highlight a specific configuration of $(\theta_\Lambda, \varphi_\Lambda)$ and  $(\theta_{\bar{\Lambda}}, \varphi_{\bar{\Lambda}})$ associated with entanglement redistribution that lies within the central detector acceptance, suggesting its potential observability.
Generating event distributions through toy Monte Carlo simulations that incorporate the maximum likelihood of the amplitude would be highly valuable, particularly for real data analysis, as it would help clarify the significance of entanglement redistribution.
For quantum discord, however, this approach would be more involved.
As noted in Eq.~\eqref{eq:clasinfor}, computing quantum discord
requires maximizing classical correlations over all local von Neumann measurements on one subsystem.
This optimization is computationally intensive even for modest event sample and demands substantial algorithmic and code-level optimization —-- a challenge that also affects future data analyses involving discord.

\begin{table}[htbp]
    \centering
    \caption{The angles $\theta_{max}$, $(\theta_\Lambda, \varphi_\Lambda)$ and  $(\theta_{\bar{\Lambda}}, \varphi_{\bar{\Lambda}})$ that realize the peak values of upper boundary of concurrence, negativity, and quantum discord for $e^+e^-\to J/\psi \to \Xi (\to \Lambda \pi)~\bar{\Xi} (\to \bar{\Lambda}\pi)$ process.}
    \label{tab:upper_xi}
    \begin{tabular}{@{}lccc@{}}
        \toprule
        $\Xi^{-}~\bar{\Xi}^{+}$ & Discord ($63.8^\circ$) & Concurrence ($66.1^\circ$) & Negativity ($69.8^\circ$)\\
        \hline
        $\theta_\Lambda$  & $284.2^\circ$ & $270.0^\circ$  & $270.0^\circ$ \\
        $\varphi_\Lambda$  & $104.2^\circ$ & $90.0^\circ$ & $90.0^\circ$\\
        $\theta_{\bar{\Lambda}}$ & $246.3^\circ$  &  $270.0^\circ$ &   $270.0^\circ$ \\
        $\varphi_{\bar{\Lambda}}$ & $123.1^\circ$  & $90.0^\circ$ &$90.0^\circ$ \\
        \hline
        $\Xi^{0}~\bar{\Xi}^{0}$ & Discord ($63.8^\circ$) & Concurrence ($62.5^\circ$) & Negativity ($66.1^\circ$)\\
        \hline
        $\theta_\Lambda$  & $246.3^\circ$ & $270.0^\circ$ & $270.0^\circ$ \\
        $\varphi_\Lambda$  & $75.8^\circ$ &$90.0^\circ$  & $90.0^\circ$\\
        $\theta_{\bar{\Lambda}}$ & $322.1^\circ$  &$270.0^\circ$  &  $270.0^\circ$  \\
        $\varphi_{\bar{\Lambda}}$ &  $47.4^\circ$ & $90.0^\circ$ &$90.0^\circ$ \\
        \hline
    \end{tabular}
\end{table}

\begin{table}[htbp]
    \centering
    \caption{The angles $\theta_{max}$, $(\theta_\Lambda, \varphi_\Lambda)$ and  $(\theta_{\bar{\Lambda}}, \varphi_{\bar{\Lambda}})$ that realize the peak values of lower boundary of concurrence, negativity, and quantum discord for $e^+e^-\to J/\psi \to \Xi (\to \Lambda \pi)~\bar{\Xi} (\to \bar{\Lambda}\pi)$ process.}
    \label{tab:lower_xi}
    \begin{tabular}{@{}lccc@{}}
        \toprule
        $\Xi^{-}~\bar{\Xi}^{+}$ & Discord ($63.8^\circ$) & Concurrence ($69.8^\circ$) & Negativity ($90.0^\circ$)\\
        \hline
        $\theta_\Lambda$  & 94.8$^\circ$ & $69.6^\circ$ &  $90.0^\circ$ \\
        $\varphi_\Lambda$  & 85.3$^\circ$ & $149.5^\circ$ & $0.0^\circ$\\
        $\theta_{\bar{\Lambda}}$ & 265.2$^\circ$  &  $ 69.1^\circ$&   $270.0^\circ$ \\
        $\varphi_{\bar{\Lambda}}$ &85.3$^\circ$  & $ 27.2 ^\circ$ &$0.0^\circ$ \\
        \hline
        $\Xi^{0}~\bar{\Xi}^{0}$ & Discord ($63.7^\circ$) & Concurrence ($69.8^\circ$) & Negativity ($90.0^\circ$)\\
        \hline
        $\theta_\Lambda$  & 94.8$^\circ$ & $72.5^\circ$ & $270.0^\circ$ \\
        $\varphi_\Lambda$  & 85.3$^\circ$ & $149.9^\circ$ & $0.0^\circ$\\
        $\theta_{\bar{\Lambda}}$ & 265.2$^\circ$ & $72.2^\circ$  &  $270.0^\circ$  \\
        $\varphi_{\bar{\Lambda}}$ & 85.3$^\circ$  & $33.7^\circ$ &$180.0^\circ$ \\
        \hline
    \end{tabular}
\end{table}

\section{Summary} \label{sec:sum}


This paper systematically investigate the complete evolution of quantum correlations in $e^+e^- \to \psi \to Y (\to \Lambda \pi)~\bar{Y} (\to \bar{\Lambda}\pi)$ processes. The concurrence, negativity, and discord of daughter system lie within a range bounded by upper and lower limits.
In comparison of mother system, entanglement of daughter system would increase in part of the phase space, the so called autodistillation in literature, and decreases in the rest, such that it is a phenomenon of entanglement redistribution in the hyperon weak decay.
It is noted that the entanglement of daughter system may decrease but will not be completely lost during the hyperon decay processes since the lower boundary of entanglement goes to zero only at scattering angle along with the beam direction.
It is pointed out that a sufficiently large transverse polarization of the hyperon is a prerequisite for an increase in entanglement following the decay of the mother particles.
This observation is particularly interesting in light of lepton beam polarization, as it can be used to control the produced hyperons' polarization \cite{Zhang:2026nwm}.
As an interesting reference, the phase-space-averaged entanglement of daughter systems consistently decreases relative to that of the mother systems.

By further exploiting discord with both extreme values and phase-space-averaged values, it is interesting that the quantumness of the daughter system is possibly increased even the entanglement characterized by concurrence and negativity is not enhanced.
This phenomenon occurs even across full phase space for $e^+e^-\to \psi(2S) \to \Xi^{0} (\to \Lambda \pi^0)~\bar{\Xi}^{0} (\to \bar{\Lambda}\pi^0)$ process, in which channel the entanglement is not increased during decay processes.
The lower boundary of the discord of daughter system indicates that there are few content of non-classical correlations within certain phase space after decay of the mother particles.

From the perspective of quantum entanglement, this work offers a new insight into the weak hadronic decays of hyperons produced in electron–positron annihilation.




\bigskip

\begin{acknowledgments}

This work is supported by the National Natural Science Foundation of China  (Grant Nos.12547111, 12447132 and 12575089) and the National Key R\&D Program of China under Grant No. 2023YFA1606703.

\end{acknowledgments}

\clearpage


\appendix

\begin{widetext}

\section{Detailed Expressions of SDM Elements}
\label{app:matrix}


The detailed expressions for ten independent elements of SDM for the daughter baryon-antibaryon pairs in Eq.~\eqref{eq:S_mu_nu_U2_matrix} are listed as follows:
\begin{align}
\Theta_{11} &= \frac{1}{2} \big(2 + \alpha_\psi + \cos2\theta(\alpha_\psi - \alpha_\Xi \alpha_{\bar{\Xi}} \cos\theta_\Lambda \cos\theta_{\bar{\Lambda}}) \nn\\
&\quad - \alpha_\Xi \alpha_{\bar{\Xi}} \cos\theta_\Lambda (\cos\theta_{\bar{\Lambda}} + 2\alpha_\psi \cos\theta_{\bar{\Lambda}} + \gamma_\psi \cos\varphi_{\bar{\Lambda}} \sin 2\theta \sin\theta_{\bar{\Lambda}}) \nn\\
&\quad + \alpha_\Xi \sin\theta_\Lambda (\alpha_{\bar{\Xi}} \cos\varphi_\Lambda (\gamma_\psi \cos\theta_{\bar{\Lambda}} \sin 2\theta + 2\cos\varphi_{\bar{\Lambda}} \sin^2\theta \sin\theta_{\bar{\Lambda}}) \nn\\
&\quad - 2\beta_\psi \cos\theta \sin\theta \sin\varphi_\Lambda)
+ 2\alpha_{\bar{\Xi}} \sin\theta \sin\theta_{\bar{\Lambda}} (\beta_\psi \cos\theta + \alpha_\Xi \alpha_\psi \sin\theta \sin\theta_\Lambda \sin\varphi_\Lambda) \sin\varphi_{\bar{\Lambda}}\big),\\
\Theta_{12} &= \frac{1}{2} \alpha_\Xi \gamma_{\bar{\Xi}} (1 + 2\alpha_\psi + \cos 2\theta) \cos\theta_\Lambda \sin\theta_{\bar{\Lambda}} - \alpha_\Xi \gamma_{\bar{\Xi}} \gamma_\psi \cos\theta \cos\varphi_\Lambda \sin\theta \sin\theta_\Lambda \sin\theta_{\bar{\Lambda}}\nn\\
&\quad + \alpha_\Xi \cos\varphi_\Lambda \sin^2\theta \sin\theta_\Lambda (\gamma_{\bar{\Xi}} \cos\theta_{\bar{\Lambda}} \cos\varphi_{\bar{\Lambda}} - \beta_{\bar{\Xi}} \sin\varphi_{\bar{\Lambda}}) \nn\\
&\quad + \alpha_\Xi \gamma_\psi \cos\theta \cos\theta_\Lambda \sin\theta (-\gamma_{\bar{\Xi}} \cos\theta_{\bar{\Lambda}} \cos\varphi_{\bar{\Lambda}} + \beta_{\bar{\Xi}} \sin\varphi_{\bar{\Lambda}}) \nn\\
&\quad + \beta_\psi \cos\theta \sin\theta (\beta_{\bar{\Xi}} \cos\varphi_{\bar{\Lambda}} + \gamma_{\bar{\Xi}} \cos\theta_{\bar{\Lambda}} \sin\varphi_{\bar{\Lambda}}) \nn\\
&\quad + \alpha_\Xi \alpha_\psi \sin^2\theta \sin\theta_\Lambda \sin\varphi_\Lambda (\beta_{\bar{\Xi}} \cos\varphi_{\bar{\Lambda}} + \gamma_{\bar{\Xi}} \cos\theta_{\bar{\Lambda}} \sin\varphi_{\bar{\Lambda}}), \\
\Theta_{13} &= -\frac{1}{2} \alpha_\Xi \beta_{\bar{\Xi}} (1 + 2\alpha_\psi + \cos 2\theta) \cos\theta_\Lambda \sin\theta_{\bar{\Lambda}} + \alpha_\Xi \beta_{\bar{\Xi}} \gamma_\psi \cos\theta \cos\varphi_\Lambda \sin\theta \sin\theta_\Lambda \sin\theta_{\bar{\Lambda}} \nn\\
&\quad + \alpha_\Xi \gamma_\psi \cos\theta \cos\theta_\Lambda \sin\theta (\beta_{\bar{\Xi}} \cos\theta_{\bar{\Lambda}} \cos\varphi_{\bar{\Lambda}} + \gamma_{\bar{\Xi}} \sin\varphi_{\bar{\Lambda}}) \nn\\
&\quad - \alpha_\Xi \cos\varphi_\Lambda \sin^2\theta \sin\theta_\Lambda (\beta_{\bar{\Xi}} \cos\theta_{\bar{\Lambda}} \cos\varphi_{\bar{\Lambda}} + \gamma_{\bar{\Xi}} \sin\varphi_{\bar{\Lambda}}) \nn\\
&\quad + \beta_\psi \cos\theta \sin\theta (\gamma_{\bar{\Xi}} \cos\varphi_{\bar{\Lambda}} - \beta_{\bar{\Xi}} \cos\theta_{\bar{\Lambda}} \sin\varphi_{\bar{\Lambda}}) \nn\\
&\quad + \alpha_\Xi \alpha_\psi \sin^2\theta \sin\theta_\Lambda \sin\varphi_\Lambda (\gamma_{\bar{\Xi}} \cos\varphi_{\bar{\Lambda}} - \beta_{\bar{\Xi}} \cos\theta_{\bar{\Lambda}} \sin\varphi_{\bar{\Lambda}}),
\\
\Theta_{14} &= \frac{1}{2} \big(\alpha_{\bar{\Xi}} (2 + \alpha_\psi) + \cos 2\theta (\alpha_{\bar{\Xi}} \alpha_\psi - \alpha_\Xi \cos\theta_\Lambda \cos\theta_{\bar{\Lambda}}) \nn\\
&\quad - \alpha_\Xi \cos\theta_\Lambda (\cos\theta_{\bar{\Lambda}} + 2\alpha_\psi \cos\theta_{\bar{\Lambda}} + \gamma_\psi \cos\varphi_{\bar{\Lambda}} \sin 2\theta \sin\theta_{\bar{\Lambda}}) \nn\\
&\quad + 2\sin\theta (\cos\theta (\alpha_\Xi \sin\theta_\Lambda (\gamma_\psi \cos\theta_{\bar{\Lambda}} \cos\varphi_\Lambda - \alpha_{\bar{\Xi}} \beta_\psi \sin\varphi_\Lambda) \nn\\
&\quad + \beta_\psi \sin\theta_{\bar{\Lambda}} \sin\varphi_{\bar{\Lambda}}) + \alpha_\Xi \sin\theta \sin\theta_\Lambda \sin\theta_{\bar{\Lambda}} (\cos\varphi_\Lambda \cos\varphi_{\bar{\Lambda}} + \alpha_\psi \sin\varphi_\Lambda \sin\varphi_{\bar{\Lambda}}))\big),\\
\Theta_{22} &= -\frac{1}{2} \gamma_\Xi \gamma_{\bar{\Xi}} (1 + 2\alpha_\psi + \cos 2\theta) \sin\theta_\Lambda \sin\theta_{\bar{\Lambda}} \nn\\
&\quad + \gamma_{\bar{\Xi}} \gamma_\psi \cos\theta \sin\theta \sin\theta_{\bar{\Lambda}} (-\gamma_\Xi \cos\theta_\Lambda \cos\varphi_\Lambda + \beta_\Xi \sin\varphi_\Lambda) \nn\\
&\quad + \gamma_\Xi \gamma_\psi \cos\theta \sin\theta \sin\theta_\Lambda (\gamma_{\bar{\Xi}} \cos\theta_{\bar{\Lambda}} \cos\varphi_{\bar{\Lambda}} - \beta_{\bar{\Xi}} \sin\varphi_{\bar{\Lambda}}) \nn\\
&\quad + \sin^2\theta (\gamma_\Xi \cos\theta_\Lambda \cos\varphi_\Lambda - \beta_\Xi \sin\varphi_\Lambda)(\gamma_{\bar{\Xi}} \cos\theta_{\bar{\Lambda}} \cos\varphi_{\bar{\Lambda}} - \beta_{\bar{\Xi}} \sin\varphi_{\bar{\Lambda}}) \nn\\
&\quad + \alpha_\psi \sin^2\theta (\beta_\Xi \cos\varphi_\Lambda + \gamma_\Xi \cos\theta_\Lambda \sin\varphi_\Lambda)(\beta_{\bar{\Xi}} \cos\varphi_{\bar{\Lambda}} + \gamma_{\bar{\Xi}} \cos\theta_{\bar{\Lambda}} \sin\varphi_{\bar{\Lambda}}),\\
\Theta_{23} &= \frac{1}{2} \beta_{\bar{\Xi}} \gamma_\Xi (1 + 2\alpha_\psi + \cos 2\theta) \sin\theta_\Lambda \sin\theta_{\bar{\Lambda}} \nn\\
&\quad + \beta_{\bar{\Xi}} \gamma_\psi \cos\theta \sin\theta \sin\theta_{\bar{\Lambda}} (\gamma_\Xi \cos\theta_\Lambda \cos\varphi_\Lambda - \beta_\Xi \sin\varphi_\Lambda) \nn\\
&\quad - \gamma_\Xi \gamma_\psi \cos\theta \sin\theta \sin\theta_\Lambda (\beta_{\bar{\Xi}} \cos\theta_{\bar{\Lambda}} \cos\varphi_{\bar{\Lambda}} + \gamma_{\bar{\Xi}} \sin\varphi_{\bar{\Lambda}}) \nn\\
&\quad - \sin^2\theta (\gamma_\Xi \cos\theta_\Lambda \cos\varphi_\Lambda - \beta_\Xi \sin\varphi_\Lambda)(\beta_{\bar{\Xi}} \cos\theta_{\bar{\Lambda}} \cos\varphi_{\bar{\Lambda}} + \gamma_{\bar{\Xi}} \sin\varphi_{\bar{\Lambda}}) \nn\\
&\quad + \alpha_\psi \sin^2\theta (\beta_\Xi \cos\varphi_\Lambda + \gamma_\Xi \cos\theta_\Lambda \sin\varphi_\Lambda)(\gamma_{\bar{\Xi}} \cos\varphi_{\bar{\Lambda}} - \beta_{\bar{\Xi}} \cos\theta_{\bar{\Lambda}} \sin\varphi_{\bar{\Lambda}}),\\
\Theta_{24} &= \frac{1}{2} \gamma_\Xi (1 + 2\alpha_\psi + \cos 2\theta) \cos\theta_{\bar{\Lambda}} \sin\theta_\Lambda \nn\\
&\quad - \cos\theta \sin\theta ((\alpha_{\bar{\Xi}} \beta_\Xi \beta_\psi - \gamma_\Xi \gamma_\psi \cos\theta_\Lambda \cos\theta_{\bar{\Lambda}}) \cos\varphi_\Lambda \nn\\
&\quad - \gamma_\Xi \gamma_\psi \cos\varphi_{\bar{\Lambda}} \sin\theta_\Lambda \sin\theta_{\bar{\Lambda}} + \alpha_{\bar{\Xi}} \beta_\psi \gamma_\Xi \cos\theta_\Lambda \sin\varphi_\Lambda \nn\\
&\quad + \beta_\Xi \gamma_\psi \cos\theta_{\bar{\Lambda}} \sin\varphi_\Lambda) + \sin^2\theta \sin\theta_{\bar{\Lambda}} (\cos\varphi_{\bar{\Lambda}} (\gamma_\Xi \cos\theta_\Lambda \cos\varphi_\Lambda - \beta_\Xi \sin\varphi_\Lambda) \nn\\
&\quad + \alpha_\psi (\beta_\Xi \cos\varphi_\Lambda + \gamma_\Xi \cos\theta_\Lambda \sin\varphi_\Lambda) \sin\varphi_{\bar{\Lambda}}),
\end{align}

\begin{align}
\Theta_{33} &= -\frac{1}{2} \beta_\Xi \beta_{\bar{\Xi}} (1 + 2\alpha_\psi + \cos 2\theta) \sin\theta_\Lambda \sin\theta_{\bar{\Lambda}} \nn\\
&\quad - \beta_{\bar{\Xi}} \gamma_\psi \cos\theta \sin\theta \sin\theta_{\bar{\Lambda}} (\beta_\Xi \cos\theta_\Lambda \cos\varphi_\Lambda + \gamma_\Xi \sin\varphi_\Lambda) \nn\\
&\quad + \beta_\Xi \gamma_\psi \cos\theta \sin\theta \sin\theta_\Lambda (\beta_{\bar{\Xi}} \cos\theta_{\bar{\Lambda}} \cos\varphi_{\bar{\Lambda}} + \gamma_{\bar{\Xi}} \sin\varphi_{\bar{\Lambda}}) \nn\\
&\quad + \sin^2\theta (\beta_\Xi \cos\theta_\Lambda \cos\varphi_\Lambda + \gamma_\Xi \sin\varphi_\Lambda)(\beta_{\bar{\Xi}} \cos\theta_{\bar{\Lambda}} \cos\varphi_{\bar{\Lambda}} + \gamma_{\bar{\Xi}} \sin\varphi_{\bar{\Lambda}}) \nn\\
&\quad + \alpha_\psi \sin^2\theta (\gamma_\Xi \cos\varphi_\Lambda - \beta_\Xi \cos\theta_\Lambda \sin\varphi_\Lambda)(\gamma_{\bar{\Xi}} \cos\varphi_{\bar{\Lambda}} - \beta_{\bar{\Xi}} \cos\theta_{\bar{\Lambda}} \sin\varphi_{\bar{\Lambda}}),\\
\Theta_{34} &= -\frac{1}{2} \beta_\Xi (1 + 2\alpha_\psi + \cos 2\theta) \cos\theta_{\bar{\Lambda}} \sin\theta_\Lambda \nn\\
&\quad - \cos\theta \sin\theta ((\alpha_{\bar{\Xi}} \beta_\psi \gamma_\Xi + \beta_\Xi \gamma_\psi \cos\theta_\Lambda \cos\theta_{\bar{\Lambda}}) \cos\varphi_\Lambda \nn\\
&\quad + \beta_\Xi \gamma_\psi \cos\varphi_{\bar{\Lambda}} \sin\theta_\Lambda \sin\theta_{\bar{\Lambda}} - \alpha_{\bar{\Xi}} \beta_\Xi \beta_\psi \cos\theta_\Lambda \sin\varphi_\Lambda \nn\\
&\quad + \gamma_\Xi \gamma_\psi \cos\theta_{\bar{\Lambda}} \sin\varphi_\Lambda) - \sin^2\theta \sin\theta_{\bar{\Lambda}} (\beta_\Xi \cos\theta_\Lambda \cos\varphi_\Lambda \cos\varphi_{\bar{\Lambda}} + \gamma_\Xi \cos\varphi_{\bar{\Lambda}} \sin\varphi_\Lambda \nn\\
&\quad - \alpha_\psi \gamma_\Xi \cos\varphi_\Lambda \sin\varphi_{\bar{\Lambda}} + \alpha_\psi \beta_\Xi \cos\theta_\Lambda \sin\varphi_\Lambda \sin\varphi_{\bar{\Lambda}}),\\
\Theta_{44} &= \frac{1}{2} \big(\alpha_\Xi \alpha_{\bar{\Xi}} (2 + \alpha_\psi) + \cos 2\theta (\alpha_\Xi \alpha_{\bar{\Xi}} \alpha_\psi - \cos\theta_\Lambda \cos\theta_{\bar{\Lambda}}) \nn\\
&\quad - \cos\theta_\Lambda (\cos\theta_{\bar{\Lambda}} + 2\alpha_\psi \cos\theta_{\bar{\Lambda}} + \gamma_\psi \cos\varphi_{\bar{\Lambda}} \sin 2\theta \sin\theta_{\bar{\Lambda}}) \nn\\
&\quad + 2\sin\theta (\cos\theta (\gamma_\psi \cos\theta_{\bar{\Lambda}} \cos\varphi_\Lambda \sin\theta_\Lambda - \alpha_{\bar{\Xi}} \beta_\psi \sin\theta_\Lambda \sin\varphi_\Lambda \nn\\
&\quad + \alpha_\Xi \beta_\psi \sin\theta_{\bar{\Lambda}} \sin\varphi_{\bar{\Lambda}}) + \sin\theta \sin\theta_\Lambda \sin\theta_{\bar{\Lambda}} (\cos\varphi_\Lambda \cos\varphi_{\bar{\Lambda}} \nn\\
&\quad + \alpha_\psi \sin\varphi_\Lambda \sin\varphi_{\bar{\Lambda}}))\big).
\end{align}
where the $\theta_\Lambda$ (or $\theta_{\bar{\Lambda}}$) is helicity angle between the $\Lambda$ (or $\bar{\Lambda}$)
momentum in the $\Xi$ (or $\bar{\Xi}$) decay plane and the $\Xi$ (or $\bar{\Xi}$) momentum
in the laboratory frame, and the $\varphi_\Lambda$ (or $\varphi_{\bar{\Lambda}}$) is the angle between the $\Xi$ (or $\bar{\Xi}$) production and decay plane.

\end{widetext}

\bibliography{entangle.bib}

\end{document}